\documentclass[aps,prb,twocolumn,floats,preprintnumbers,amsmath,amssymb]{revtex4}

\usepackage{graphicx}
\usepackage{dcolumn}
\usepackage{bm}

\begin{document}

\preprint{Catal. Lett. (submitted)}
\title{First-Principles Approach to Heat and Mass Transfer Effects\\ in Model Catalyst Studies}

\author{Sebastian Matera}
\email{matera@fhi-berlin.mpg.de}
\affiliation{Fritz-Haber-Institut der Max-Planck-Gesellschaft, Faradayweg 4-6, D-14195 Berlin, Germany}

\author{Karsten Reuter}
\email{reuter@fhi-berlin.mpg.de}
\affiliation{Fritz-Haber-Institut der Max-Planck-Gesellschaft, Faradayweg 4-6, D-14195 Berlin, Germany}

\received{August 21, 2009}

\begin{abstract}
We assess heat and mass transfer limitations in {\em in situ} studies of model catalysts with a first-principles based multiscale modeling approach that integrates a detailed description of the surface reaction chemistry and the macro-scale flow structures. Using the CO oxidation at RuO$_2$(110) as a prototypical example we demonstrate that factors like a suppressed heat conduction at the backside of the thin single-crystal, and the build-up of a product boundary layer above the flat-faced surface play a significant role.
\end{abstract}

\maketitle

\section{Introduction}

During practical operation heterogeneous catalysts are exposed to rather harsh environments with reactant partial pressures of the order of atmospheres and quite elevated temperatures. Quite in contrast to these conditions, much of our atomic-scale understanding of the catalytic function derives to date from controlled experiments of single-crystals in ultra-high vacuum (UHV). In order to bridge particularly the resulting {\em pressure gap} much progress has recently been achieved in pushing {\em in-situ} techniques that are capable of delivering equally resolved and quantitative information for such model catalysts in technologically relevant gas-phases \cite{1}. The focus in such studies is on identifying possible differences in the surface chemistry at corresponding near-ambient pressures and elevated temperatures. However, one also has to recognize that at then achievable, much higher conversion rates heat and mass transfer effects in the gas-surface system become increasingly 
important. In contrast to low-conversion operation in UHV this concerns for instance the gas-phase transport of formed products away from the active surface, and how efficiently the large amount of heat generated by the exothermic surface reactions can dissipate in the system.

Differences with respect to these factors in present experimental {\em in-situ} setups may possibly account for some of the existing controversies in the field. Similarly, corresponding effects are important for comparisons with quantitative data from emerging first-principles based microkinetic modeling approaches \cite{2}. Among these, latest first-principles kinetic Monte Carlo (1p-kMC) modeling even provides a microscopically correct account of the spatial arrangement and interactions of the adsorbed chemicals \cite{3}. The thereby achieved predictive-quality description of the detailed surface reaction chemistry must then be complemented by an appropriate treatment of the macroscopic heat and mass transfer in the gas-surface system. Here we aim to qualify the latter effects precisely for the near-ambient, high-activity conditions targeted by the novel {\em in-situ} techniques. We therefore present an approach which efficiently integrates accurate 1p-kMC simulations for the surface kinetics into computational fluid dynamics (CFD) simulations. Using the CO oxidation at RuO$_2$(110) as a prototypical example, we illustrate that the peculiarities of the thin single-crystal reactor geometry can readily lead to heat dissipation and mass transport limitations that severely affect the observable catalytic function. If these limitations are not appropriately accounted for in both experiment and theory, wrong mechanistic conclusions about the atomic-scale surface kinetics at technologically relevant gas-phase conditions may easily arise, thereby blocking the envisioned route towards a rational design of future improved catalysts.

\section{Methods}

\begin{figure}
 \includegraphics[width=8cm]{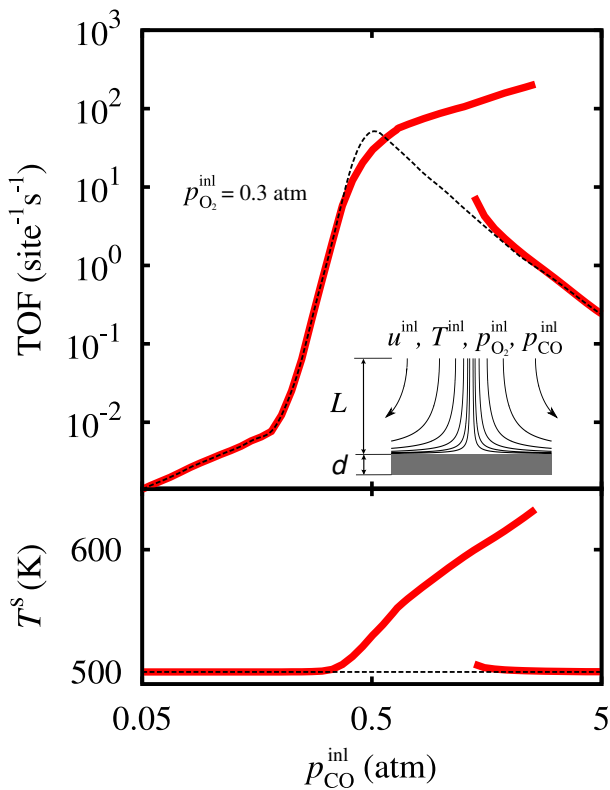}
  \caption{Comparison of intrinsic turnover frequencies (TOFs, black dashed line) with observable TOFs (red solid line) when accounting for heat and mass transfer effects. In the considered stagnation flow geometry the inlet is placed at distance $L$ = 1 cm above the RuO$_2$(110) surface as schematically explained in the inset. Shown is steady-state data for $T^\text{inl}$ = 500K, $p^\text{inl}_\text{O$_2$}$ = 0.3 atm, $u^\text{inl}$ = 1 cm/s and varying $p^\text{inl}_\text{CO}$ at the inlet. Due to the suppressed heat flux at the back of the $d$ = 1mm thin single-crystal the system is able to sustain a high-activity branch for more CO-rich conditions than the nominal most active ``state'', where the surface temperature, $T^\text{s}$ shown in the lower panel, is significantly increased.}
\label{figure1}
\end{figure}

At the continuum level a reliable description of the macro-scale flow structures is achieved by the transient Navier-Stokes equations together with energy and species governing equations \cite{4}. Aiming for a general assessment of heat and mass transfer limitations we confine the attention to the center of a flat-faced single-crystal model catalyst of thickness $d$ and consider the axisymmetric stagnation flow geometry shown in Fig. \ref{figure1}. In this geometry \cite{4}, the equations are closed by appropriate boundary conditions at the inlet at a macroscopic distance $z = L$ and at the solid surface at $z = 0$. For the inlet this is obviously the control of the temperature, density and composition of the gas mixture, here $T^\text{inl}$, $p^\text{inl}_\text{O$_2$}$ , $p^\text{inl}_\text{CO}$ and $p^\text{inl}_\text{CO$_2$}$, as well as its axial velocity $u^\text{inl}$. With thus defined inlet conditions, we here specifically target stationary operation, i.e. the steady-state catalytic conversion rates, so-called turnover frequencies (TOFs, in units of product molecules per surface active site and time). In this case, the partial mass fluxes of reactants and products at the gas-solid interface cancel to yield a total mass flux of zero. For near-ambient pressures we can furthermore assume the temperature to be continuous across the interface. The surface boundary conditions that then remain to be determined are the species conversion rate and connected heat release through the on-going surface chemical reactions, as well as the heat flux into the solid. 

For the prior two quantities we rely on the accurate first-principles description provided by 1p-kMC simulations, which evaluate the statistical interplay between the elementary processes involved in the catalytic cycle \cite{3}. In the established model of CO oxidation at RuO$_2$(110) this is specifically the set of 26 elementary processes defined by all non-correlated site and element specific adsorption, desorption, diffusion and reaction events that can occur on a lattice spanned by two different active sites offered by the surface \cite{5,6}. For a given gas-phase impingement as described by a local temperature $T^\text{s}$ and reactant partial pressures $p^\text{s}_\text{O$_2$}$ and $p^\text{s}_\text{CO}$ directly at the surface, these simulations provide the average rate of reactant and product adsorption, desorption and conversion per surface area and time, i.e. the partial mass fluxes at the surface required for the CFD modeling. Multiplying the conversion rate with the specific enthalpy change connected with one CO + O $\rightarrow$ CO$_2$ reaction event provides furthermore the released heat rate that enters into the heat balancing equation. While it is only these average fluxes and heat rate that matter for the macroscopically described flow field, it is important to note that they are still properly derived from microscopic simulations that fully account for the site heterogeneity and distributions at the surface. This is thus distinctly different to mean-field based phenomenological descriptions that are commonly integrated in the CFD modelling of macro-scale flow structures \cite{7} and which have for the CO oxidation at RuO2(110) been shown to fail qualitatively \cite{8}.

The also required heat flux into the solid is determined by the heat transport inside the crystal of width $d$ (described through the bulk heat conductivity) and the degree of heat dissipation that is possible at the back of the crystal, e.g. through radiative loss or contact with the sample holder. With real experimental setups lying anywhere in between we analyse the relevance of this factor for thin single-crystals by considering two opposite extremes: Fixed temperature at the sample backside to mimic a highly efficient heat coupling of the crystal to the system, and zero heat flux at the sample backside to represent a well insulated sample.

\section{Results and Discussion}

In the adiabatic situation with a suppressed heat flux through the back of the sample the main dissipation channel left for the heat released by the exothermic surface reactions is into the surrounding gas-phase. Above a critical value of the TOF (and therewith generated heat rate) this channel may no longer be efficient enough to maintain the nominal surface temperature. If increasing the surface temperature $T^\text{s}$ then furthermore enables higher conversion rates, the system can run away into a new highly active steady-state that is characterized by sizable temperature gradients across the system. Considering representative parameters for the crystal width $d$ = 1 mm, inlet distance $L$ = 1 cm and axial inlet flow velocity $u^\text{inl}$ = 1 cm/s in our simulations for the CO oxidation at RuO$_2$(110) we find this critical TOF to be of the order of 10 site$^{-1}$ sec$^{-1}$ at near-ambient partial pressures. Variations of $d$, $L$ and $u^\text{inl}$ over one order of magnitude change this value by a similar amount, but leave the qualitative physics discussed in the following untouched. 

In the most active ``state'' intrinsic steady-state conversions above this critical TOF number are reached for temperatures above about 500 K, and Fig. \ref{figure1} illustrates the sizable effect that the surface heat-up already has on the observable activity at this threshold. Compared are the intrinsic steady-state TOFs as resulting from the 1p-kMC simulations, i.e. assuming that the temperature and partial pressures at the surface are identical to those at the inlet, and the really observable steady-state TOFs when explicitly accounting for heat and mass transfer effects through the coupled 1p-kMC+CFD approach. Shown is a set of gas-phase conditions for fixed $p^\text{inl}_\text{O$_2$}$ = 0.3 atm that comprises with increasing $p^\text{inl}_\text{CO}$ the three characteristic ``states'' of the surface: O-poisoned at the lowest $p^\text{inl}_\text{CO}$ shown, CO-poisoned at the highest  $p^\text{inl}_\text{CO}$ shown, and in between the most active ``state'' with both reactants coexisting in appreciable amounts at the surface \cite{7}. For a range of more CO-rich conditions than this intrinsically most active ``state'' of the surface, the 1p-kMC+CFD results in Fig. \ref{figure1} reveal that the system is in fact able 
to sustain a high-activity steady-state operation, in which the surface temperature is up to 150 K higher than the nominal inlet temperature. This increased $T^\text{s}$ is the result of the run away heat-up and only remains at a finite steady-state value because further heating would lower the surface catalytic activity; for the near-ambient conditions shown primarily because enhanced CO desorption would deplete the CO population at the surface. This limitation by CO depletion also rationalizes why a high-activity branch only exists for more CO-rich conditions than the intrinsically most active ``state'' towards the right in Fig. \ref{figure1}, and why the corresponding steady-state surface temperature progressively increases with increasing $p^\text{inl}_\text{CO}$. As a net result, the observable TOF profile shown in Fig. \ref{figure1} is significantly changed from the intrinsic one, with the highest absolute rates about one order of magnitude higher and at a partial pressure ratio that is significantly shifted towards more CO-rich conditions.

\begin{figure}
 \includegraphics[width=8cm]{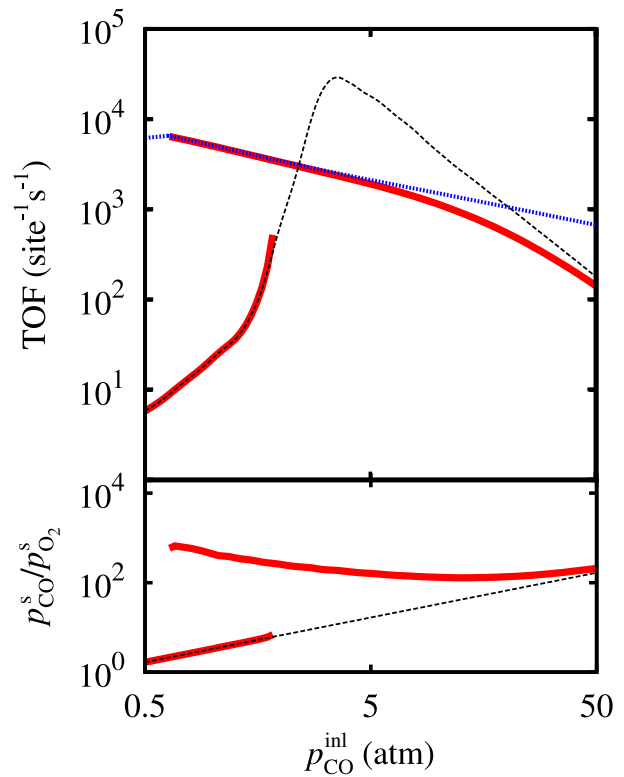}
  \caption{Same as Fig. \ref{figure1}, but for an ideal heat coupling at the backside of the crystal. With all other parameters identical to those in Fig. \ref{figure1}, the significantly increased intrinsic TOFs (black dashed line) at $T^\text{inl}$ = 600K now induce an extended product boundary layer above the flat-faced model catalyst with the concomitant mass transfer limitations affecting the observable TOF profile (red solid line). For a wide range of the gas-phase conditions shown the observable TOF profile coincides in fact with the upper limit set purely by the gas-phase mass transport (blue dotted line), i.e. with a limit that is completely independent of the actual model catalyst employed. The lower panel shows the corresponding partial pressure ratio directly at the surface and illustrates that in the high-activity branch, the surface effectively sees a more CO-rich gas phase compared to the nominal inlet composition.}  \label{figure2}
\end{figure}

The pronounced surface temperature increase is a direct consequence of the adiabatic boundary at the back of the single-crystal and demonstrates the relevance of the detailed heat transfer through the substrate on the {\em in-situ} measurable activity. If we move to the opposite extreme where an ideal heat coupling is able to maintain the nominal temperature at the sample backside, we never obtain any significant increase of $T^\text{s}$ even when going to higher temperatures around 600\,K, where the intrinsic TOFs in the most active ``state'' exceed 10$^4$ site$^{-1}$sec$^{-1}$.\cite{5,6} However, at such high conversions an extended product boundary layer rapidly builds up above the flat-faced single-crystal surface with concomitant mass transfer limitations then affecting the observable TOFs. This is illustrated in Fig. \ref{figure2}, where at $T^\text{inl}$ = 600 K the observable TOFs are never as high as in the intrinsic most active ``state'', but with a high-activity steady-state branch now extending to much less CO-rich conditions. The latter is possible because the boundary layer limits the transport of both diatomic reactants in a similar way. As long as the nominal inlet composition is more CO-rich than stoichiometric feed this effectively increases the $p^\text{s}_\text{CO} / p^\text{s}_\text{O$_2$}$ ratio directly at the surface as shown in the lower panel of Fig. \ref{figure2} and thereby creates an impingement that brings the surface closer to its most active ``state''.

Even in this ideal isothermal case the true catalytic function is thus masked, this time by mass transfer limitations at the flat-faced surface. In fact, as also shown in Fig. \ref{figure2} the observable TOF is under these conditions very close to the upper limit set by pure gas-phase mass transfer. This limit is completely independent of the actual model catalyst employed, and as apparent from Fig. \ref{figure2} for a wide range of gas-phase conditions it is significantly lower than the intrinsic catalytic activity of the RuO$_2$(110) surface. With the single-crystal in real experimental setups neither perfectly heat-coupled nor isolated, the two effects discussed here separately in Figs. \ref{figure1} and \ref{figure2} will obviously be intricately intermingled and need to be disentangled by dedicated measurements and setups. This holds even more as the high-activity branches in Figs. \ref{figure1} and \ref{figure2} are not the only steady-state solution. Instead there is in both cases a range of gas-phase conditions, in which the system can also operate in a low-activity mode that never exceeds the critical TOF and therefore coincides with the intrinsic activity. While the intrinsic surface kinetics of the employed 1p-kMC model does not provide multiple steady-states \cite{7}, they thus enter at larger length scales through the coupling to the surrounding flow field. A corresponding bi-stability clearly suggests that the system could oscillate between the two modes, possibly even inhomogeneously in form of reaction fronts over the single-crystal surface. In case of the first discussed heat transfer limitations, an intuitive propagation mechanism would hereby be via the formation of local hot spots, while in the latter discussed mass transfer case it would be via gas-phase coupling \cite{9}, with the presented approach establishing the intriguing possibility to quantify these model conceptions with first-principles based simulations.

\section{Conclusions}

We have presented a first-principles based multiscale modeling approach to heterogeneous catalysis integrating accurate 1p-kMC 
simulations for the surface kinetics into CFD modeling of the macro-scale flow structures. While its efficient formulation readily allows addressing more complex reactor geometries, we have used it here to specifically assess heat and mass transfer effects in {\em in-situ} studies of single-crystal model catalysts. Using RuO$_2$(110) as a representative substrate for the frequently studied CO oxidation reaction we find corresponding limitations to significantly mask the intrinsic catalytic function at the high conversion rates reached at near-ambient gas-phase conditions. Two crucial and hitherto largely unappreciated factors specific to the model catalyst reactor geometry are in this respect the degree of heat dissipation at the back of the thin sample and the propensity to build-up a product boundary layer above the flat-faced surface with its high density of active sites spread over a macroscopic area. A qualified discussion of a possibly different surface chemistry across the {\textquotedblleft}pressure gap{\textquotedblright} requires accounting for resulting temperature and pressure gradients, both in experimental {\em in-situ} setups and through predictive-quality integrated models as the one presented here. Otherwise wrong mechanistic conclusions may be derived which hamper our progress towards an atomic-scale understanding of the function of heterogeneous catalysts at technologically relevant gas-phase conditions.

\section*{Acknowledgements}
Funding within the MPG Innovation Initiative {\em Multiscale Materials Modeling of Condensed Matter} and the DFG Cluster of Excellence {\em Unifying Concepts in Catalysis} is gratefully acknowledged.

\end{document}